\documentclass[submitting]{nst}

\usepackage{amsmath}
\usepackage{graphicx}
\usepackage{subcaption}
\usepackage{booktabs}
\usepackage{placeins}
\usepackage{xcolor}
\usepackage{siunitx}
\usepackage{multirow}
\DeclareSIUnit{\standardliter}{SL}
\DeclareSIUnit{\ppb}{ppb}
\DeclareSIUnit{\PE}{PE}
\renewcommand{\linenumbers}{}

\newcommand{\tsinghuaphy}{\affiliation{Department of Physics \& Center for High Energy Physics, Tsinghua University, Beijing 100084, China}}
\newcommand{\tsinghuaephy}{\affiliation{Key Laboratory of Particle and Radiation Imaging (Ministry of Education) \& Department of Engineering Physics, Tsinghua University, Beijing 100084, China}}
\newcommand{\ustc}{\affiliation{State Key Laboratory of Particle Detection and Electronics, University of Science and Technology of China, Hefei 230026, China}}
\newcommand{\ustcimp}{\affiliation{
		Department of Modern Physics, University of Science and Technology of China, Hefei 230026, China}}
\newcommand{\westlake}{\affiliation{School of Science, Westlake University, Hangzhou 310030, China}}
\newcommand{\sysuphy}{\affiliation{School of Physics, Sun Yat-sen University, Guangzhou 510275, China}}
\newcommand{\sysuifcen}{\affiliation{Sino-French Institute of Nuclear Engineering and Technology, Sun Yat-sen University, Zhuhai 519082, China}}
\newcommand{\cuhk}{\affiliation{School of Science and Engineering, The Chinese University of Hong Kong (Shenzhen), Shenzhen, Guangdong, 518172, China}}
\newcommand{\buaasp}{\affiliation{School of Physics, Beihang University, Beijing 100083, China}}
\newcommand{\buaalab}{\affiliation{Beijing Key Laboratory of Advanced Nuclear Materials and Physics, Beihang University, Beijing 100191, China}}

\begin{document}

\title{Oxygen Outgassing Measurements and Mitigation for the RELICS Liquid Xenon Time Projection Chamber}
\author{Jingfan Gu}\tsinghuaphy
\author{Jiachen Yu}\ustc\ustcimp

\author{Chang Cai}\tsinghuaphy
\author{Jiangyu Chen}\sysuifcen
\author{Huayu Dai}\cuhk
\author{Rundong Fang}\buaasp
\author{Fei Gao}\tsinghuaphy
\author{Hongrui Gao}\tsinghuaphy
\author{Jiheng Guo}\buaasp
\author{Xiaoran Guo}\ustc\ustcimp
\author{Xu Han}\tsinghuaphy
\author{Yanzhou Hao}\tsinghuaphy
\author{Zhen Hu}\tsinghuaphy
\author{Yang Lei}\tsinghuaphy
\author{Kaihang Li}\thanks{\emph{Corresponding author: } likh23@mails.tsinghua.edu.cn} \tsinghuaphy
\author{Shengchao Li}\westlake
\author{Siyin Li}\westlake
\author{Yuechen Li}\cuhk
\author{Qing Lin}\ustc\ustcimp
\author{Jiajun Liu}\sysuphy
\author{Zhu Liu}\westlake
\author{Yuanyuan Ren}\cuhk
\author{Lijun Tong}\ustc\ustcimp
\author{Yuhuang Wan}\tsinghuaphy
\author{Jun Wang}\westlake
\author{Wei Wang}\sysuphy\sysuifcen
\author{Xiaoping Wang}\buaasp\buaalab
\author{Xiaoyu Wang}\westlake
\author{Yuehuan Wei}\sysuifcen
\author{Xiang Xiao}\sysuphy
\author{Lingfeng Xie}\tsinghuaphy
\author{Jijun Yang}\westlake
\author{Litao Yang}\tsinghuaephy
\author{Jingqiang Ye}\cuhk
\author{Qian Yue}\tsinghuaephy
\author{Yuyong Yue}\sysuifcen
\author{Tianyuan Zha}\cuhk
\author{Honghui Zhang}\cuhk
\author{Zhicai Zhang}\tsinghuaephy
\author{Yifei Zhao}\tsinghuaphy

\collaboration{RELICS Collaboration}
\thanks{\emph{Collaboration email: } relics@tsinghua.edu.cn}
\noaffiliation

\begin{abstract}
The REactor neutrino LIquid xenon Coherent elastic Scattering (RELICS) experiment searches for reactor coherent elastic neutrino-nucleus scattering (CE$\nu$NS) using a dual-phase liquid xenon time projection chamber, where an ionization-only analysis requires a long electron lifetime. Material outgassing is a persistent source of electronegative impurities and must be quantified for reliable purity predictions and suppressed to improve xenon purity. In this work, we developed a rate-of-rise system combined with residual gas analysis to measure oxygen-specific outgassing rates of detector materials. Their temperature dependence was studied to estimate cryogenic impurity loads. To test whether material outgassing can account for the prototype impurity load, these estimates were compared with source terms inferred from fits of the RELICS purity-evolution model to prototype data. PTFE is an important contributor to the estimated detector impurity load. A vacuum-baking treatment was therefore evaluated and shown to substantially reduce its outgassing, although subsequent air exposure weakened this benefit. These results provide material-source inputs for predicting the purity evolution of future RELICS detectors.
\end{abstract}

\keywords{Reactor CE$\nu$NS, liquid xenon time projection chamber, material outgassing, vacuum baking}

\maketitle

\section{Introduction}
\label{sec:intro}

Coherent elastic neutrino-nucleus scattering (CE$\nu$NS) is a neutral-current process in the Standard Model, in which a low-momentum-transfer neutrino scatters coherently from an entire nucleus~\cite{Freedman1974}. Owing to the coherent enhancement of the cross section, approximately proportional to the square of the neutron number, CE$\nu$NS provides a sensitive probe of electroweak interactions~\cite{canas2018future}, nuclear structure~\cite{patton2012neutrino}, astrophysical neutrino processes~\cite{beacom2010diffuse, Drukier1984}, and possible physics beyond the Standard Model~\cite{coloma2017curtailing}. Since the first observation of CE$\nu$NS by the COHERENT collaboration using a cesium iodide (CsI) detector~\cite{COHERENT_CsI}, the process has been investigated across various target materials and technologies. The COHERENT collaboration subsequently reported measurements with a liquid argon target~\cite{COHERENT_LAr}, and the CONUS+ experiment achieved the first observation of reactor antineutrino CE$\nu$NS using ultra-low-threshold high-purity germanium detectors~\cite{CONUSplus}. Additionally, CE$\nu$NS signals from solar $^{8}\mathrm{B}$ neutrinos have been measured in multi-ton dual-phase liquid xenon time projection chambers by the XENONnT~\cite{XENONnT_CEvNS}, PandaX-4T~\cite{PandaX_CEvNS}, and LZ~\cite{LZ_CEvNS} collaborations. Despite these advances, reactor-neutrino CE$\nu$NS remains an important experimental frontier, where independent measurements with different target nuclei and low-threshold detector technologies are essential for signal confirmation and for controlling systematic uncertainties.

The RELICS (REactor neutrino LIquid xenon Coherent elastic Scattering) experiment is being developed to measure reactor-neutrino CE$\nu$NS at the Sanmen Nuclear Power Plant using a dual-phase liquid xenon time projection chamber (LXeTPC) ~\cite{cai2024reactor}. In a conventional dual-phase LXeTPC, an energy deposition produces prompt scintillation light (S1) and ionization electrons, the latter of which drift upward and generate a delayed ionization signal (S2) through electroluminescence in the gas phase. For reactor CE$\nu$NS, however, the expected nuclear recoil energies are in the keV or sub-keV range. In this regime, the S1 signal can fall below the detection threshold, making an S2-only analysis necessary for lowering the effective energy threshold and improving sensitivity to small ionization signals. As a result, the performance of RELICS depends critically on the survival of drifting electrons and therefore on the purity of the liquid xenon.

Electronegative impurities in liquid xenon can capture drifting electrons and shorten the electron lifetime, $\tau_e$, which is the key purity parameter for the S2-only analysis in RELICS. To describe this effect at the detector level, a purity-evolution model has been developed for RELICS and validated using RELICS prototype data obtained under different circulation and purification conditions~\cite{yu2026projection, xie2026development}. In this model, material outgassing enters as an effective impurity source term that competes with getter purification and determines the long-term equilibrium purity. Therefore, independent measurements of material outgassing rates, together with methods to suppress the dominant sources, are necessary for establishing a reliable purity-control strategy for RELICS.

Based on this motivation, this paper presents a systematic study of material outgassing and its suppression for the RELICS LXeTPC. Oxygen-specific outgassing rates are measured for the main non-metallic detector materials and selected detector components, including polytetrafluoroethylene (PTFE, commercially known as Teflon), Kapton, polyether ether ketone (PEEK), and the photomultiplier tube (PMT) base, using an accumulation method combined with a residual gas analyzer (RGA). The temperature dependence of the material outgassing rates is then investigated to estimate the impurity load under cryogenic operating conditions. The resulting cryogenic outgassing rates are scaled to the detector geometry and compared with the outgassing source term required by the RELICS purity-evolution model. Based on these results, an outgassing-suppression method is developed and evaluated for reducing the dominant impurity contribution. Combined with the purification model, these measurements provide the material-source inputs needed to predict the purity evolution of future RELICS detectors.

\section{Material outgassing characterization}

In long-term LXeTPC operation, outgassing from internal detector components is one of the major sources of electronegative impurities. We therefore performed a systematic study of the outgassing rates of the materials used in the detector, with the goal of identifying the dominant contamination sources and developing effective suppression methods for achieving long electron lifetimes.

To interpret the measured outgassing rates and relate vacuum measurements to detector operation, we first consider the physical mechanism of material outgassing. For a bulk material, outgassing can be described as a two-step process: gas molecules absorbed in the material diffuse toward the surface and then desorb into the surrounding environment. The overall rate is controlled by the slower of these two processes. In the RELICS TPC, electronegative impurities released from detector materials are continuously transported away by the circulating xenon and removed by the purification system. The impurity concentration near the material surface can therefore be approximated as low. Under this condition, we treat molecules reaching the surface as being rapidly released and removed, so that the outgassing process is mainly limited by diffusion through the material.

A similar boundary condition is realized in high-vacuum measurements, where continuous pumping maintains a low impurity partial pressure near the material surface~\cite{duncan2005permeation}. Vacuum measurements provide a practical proxy for comparing the outgassing behavior of detector materials and for estimating the impurity source terms needed in the purity-evolution model. The diffusion model, measurement setup, and material outgassing results are described in the following sections.

\subsection{Diffusion-limited outgassing model}

Following the diffusion-limited picture introduced above, we describe the time evolution of the gas concentration inside a bulk material using Fick's second law of diffusion~\cite{Sammartano_P35}:

\begin{equation}
    \frac{\partial C(x,t)}{\partial t} = D \frac{\partial^2 C(x,t)}{\partial x^2}
    \label{eq:ficks_law}
\end{equation}
where $C(x,t)$ denotes the pressure-equivalent gas concentration (Pa) inside the material at a depth $x$ and time $t$, and $D$ is the material-dependent diffusion coefficient with units of $\mathrm{m^2\,s^{-1}}$.

For an idealized slab with thickness $2d$ and uniform initial gas concentration $C_0$, both surfaces are treated as a perfect sink under high-vacuum pumping. The corresponding initial and     boundary conditions are
\begin{align}
    C(x,0) &= C_0 \quad (0<x<2d), \label{eq:initial_cond} \\
    C(0,t) &= C(2d,t)=0 \quad (t>0). \label{eq:boundary_cond}
\end{align}
Solving Eq.~\eqref{eq:ficks_law} under these conditions gives the specific outgassing rate, which is obtained from the concentration gradient at the surface according to Fick's first law~\cite{OHanlon_Vacuum_Tech}:
\begin{equation}
    \Lambda(t)
    =
    \frac{2DC_0}{d}
    \sum_{n=0}^{\infty}
    \exp\left[
    -\frac{(2n+1)^2\pi^2Dt}{4d^2}
    \right],
    \label{eq:outgassing_full}
\end{equation}
where $\Lambda(t)$ represents the gas load released per unit surface area per unit time with units of $\mathrm{Pa\cdot m\cdot s^{-1}}$.

The full solution admits two useful limiting forms. At short times, $t\ll d^2/D$, the diffusion length is much smaller than the half-thickness of the slab, and the slab can be approximated as a semi-infinite medium~\cite{Sammartano_P35,OHanlon_Vacuum_Tech,chiggiato2020outgassing}. In this limit,
\begin{equation}
\Lambda(t)
\approx
C_0\left(\frac{D}{\pi t}\right)^{1/2}.
\label{eq:outgassing_short}
\end{equation}
At long times, $t\gg d^2/D$, the higher-order terms decay rapidly, and the first term of the series dominates, so the outgassing rate reduces to ~\cite{OHanlon_Vacuum_Tech,chiggiato2020outgassing}
\begin{equation}
\Lambda(t)
\approx
\frac{2DC_0}{d}
\exp\left(
-\frac{\pi^2Dt}{4d^2}
\right).
\label{eq:outgassing_long}
\end{equation}
Therefore, under the present assumptions, the diffusion-limited model predicts a transition from an early-time power-law decay to a long-time exponential decay.

The boundary between these two regimes is determined by the characteristic diffusion time scale $d^2/D$, which depends on both the sample geometry and the material-dependent diffusion coefficient. Since the tested samples have different effective thicknesses and unknown diffusion coefficients, the experimental time window cannot be assigned to a universal short-time or long-time regime for all materials. In addition, real materials may deviate from the ideal slab model because of surface effects, material inhomogeneity, non-uniform initial gas loading, and multiple outgassing species. Therefore, to provide a consistent description of the measured outgassing behavior, we employ phenomenological models that are selected according to the observed decay characteristics of each material. Depending on the experimental data, the outgassing rate may be described by a generalized power-law function, an exponential function, or piecewise combinations of these functions. Similar phenomenological descriptions have been widely adopted in vacuum outgassing studies of engineering materials, including both near-diffusion-limited $t^{-1/2}$ behavior and generalized power-law decays with non-integer exponents~\cite{Huang2016outgassing}.

When the measured outgassing rate exhibits a continuous power-law decay over the experimental time window, it is fitted using
\begin{equation}
\Lambda(t)=k t^{-\alpha},
\label{eq:outgassing_fit}
\end{equation}
where $k$ is the normalization factor and $\alpha$ is the decay exponent. 

For materials whose outgassing rate exhibits an exponential decay over the measured time window, the data are fitted using
\begin{equation}
\Lambda(t)=A\exp(-t/\tau),
\label{eq:outgassing_fit_e}
\end{equation}
where $A$ is the normalization factor and $\tau$ is the characteristic time.

For some materials, the outgassing curve exhibits a clear transition during the measurement. In such cases, piecewise models are adopted to describe the different decay regimes.

The fitted parameters provide a phenomenological description of the measured outgassing behavior and enable a consistent comparison among different materials.

\subsection{Room-temperature outgassing measurement system and method}

\begin{figure}[htbp]
    \nolinenumbers
    \centering
    \includegraphics[width=0.4\textwidth]{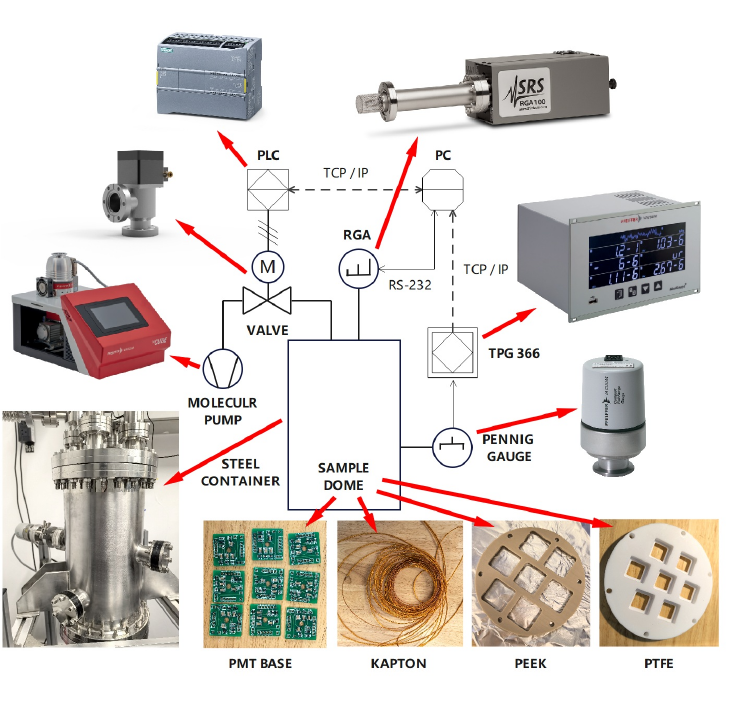}
    \captionsetup{justification=raggedright,singlelinecheck=false}
    \caption{
    Schematic of the room-temperature outgassing measurement system based on the accumulation method. 
    The system consists of a stainless-steel sample chamber, a high-vacuum pumping line, a Penning gauge, and an RGA for oxygen partial-pressure measurements.}
    \label{fig:outgassing_system}
\end{figure}

An automated room-temperature outgassing measurement system was developed to measure material outgassing rates, as shown schematically in Fig.~\ref{fig:outgassing_system}. The system consists of a stainless-steel sample chamber, a high-vacuum pumping line (HiCube 80 Eco, PFEIFFER), a PLC-controlled pneumatic valve, vacuum gauges (PKR 251, PFEIFFER), an RGA (RGA100, Stanford Research System), and a PC-based data acquisition system. The sample chamber is connected to the pumping line through the PLC-controlled pneumatic valve and is evacuated by a molecular pump. The total pressure is monitored by the vacuum gauges through a TPG366 controller, while the RGA is used to monitor the partial pressures of residual gas species.

The measurement procedure consists of two parts: the measurement of the time-dependent total-gas load from the chamber and its contents, and the determination of the oxygen partial-pressure fraction. These quantities are combined to obtain the oxygen gas load before normalization by the test-sample quantity.

The total-gas load is measured using the accumulation method~\cite{chiggiato2020outgassing}. During each measurement cycle, the sample chamber is isolated from the pumping line by closing the pneumatic valve. The pressure increase in the isolated chamber is caused by gas released from the sample inside, together with the residual background contribution from the chamber. The initial linear pressure-rise rate is defined as
\begin{equation}
    s_{\text{total}}(t)
    =
    \frac{dP_{\text{total}}}{d\tau},
    \label{eq:pressure_slope_total}
\end{equation}
where $t$ denotes the elapsed pumping time and $\tau$ denotes the elapsed time after valve closure within an individual accumulation cycle. The derivative is evaluated over the initial linear pressure-rise interval, and $s_{\text{total}}(t)$ has units of $\mathrm{Pa\,s^{-1}}$. The corresponding physical total-gas load is
\begin{equation}
    Q_{\text{total}}^{\mathrm{phys}}(t)
    =
    V s_{\text{total}}(t),
    \label{eq:outgassing_total_load}
\end{equation}
where $V$ is the sample-chamber volume expressed in liters, so that $Q_{\text{total}}^{\mathrm{phys}}(t)$ has units of $\mathrm{Pa\,L\,s^{-1}}$. This is the load from the chamber and its contents before any background treatment or sample normalization; no ppb scaling is applied to the measured pressure slope.

Since oxygen is the main electronegative impurity affecting the electron drift lifetime, the oxygen fraction in the released gas is determined in a separate RGA measurement. Continuous operation over the multi-day measurement is avoided because the chamber pressure may exceed the RGA operating range and prolonged operation would shorten the lifetime of the filament and electron multiplier.

In this separate measurement, the chamber is first pumped down to the operating pressure required for the RGA. The RGA is then operated in leak-test mode to monitor the $\text{O}_2$ partial pressure at $m/z=32$, while the vacuum gauge simultaneously records the total pressure. After the pneumatic valve is closed, the oxygen partial-pressure rise and the total-pressure rise are measured over the same time interval. Their initial slopes are denoted as $s_{\text{O}_2}$ and $s_{\text{total}}$, respectively. Assuming that the gas composition remains constant during the measurement interval, the ratio of the pressure-rise rates equals the oxygen fraction in the released gas, so the oxygen fraction is estimated as
\begin{equation}
    f_{\text{O}_2}
    =
    \frac{s_{\text{O}_2}}{s_{\text{total}}}.
    \label{eq:oxygen_fraction}
\end{equation}


The measured oxygen fraction is assumed to remain constant throughout the multi-day total-gas measurement. The physical oxygen gas load is obtained by multiplying the total-gas load by this fraction:
\begin{equation}
    Q_{\text{O}_2}^{\mathrm{phys}}(t)
    =
    f_{\text{O}_2}Q_{\text{total}}^{\mathrm{phys}}(t).
    \label{eq:oxygen_specific_rate}
\end{equation}
For reporting in the ppb convention used by the purity-evolution model, the numerical oxygen loads are multiplied by $10^9$:
\begin{equation}
    q_{\text{O}_2}(t)=10^9 Q_{\text{O}_2}^{\mathrm{phys}}(t).
    \label{eq:scaled_oxygen_load}
\end{equation}
The scaled loads $q$ are quoted in $\mathrm{ppb\,Pa\,L\,s^{-1}}$; the factor $10^9$ is a reporting conversion and does not represent an additional physical gas load. The empty-chamber load and its asymptotic background term follow the same convention. The room-temperature material curves and their fit amplitudes refer to the complete test sample. The corresponding rate per unit sample quantity is defined as
\begin{equation}
    \Lambda_{m,\mathrm{O}_2}^{\mathrm{RT}}(t)
    =\frac{q_{m,\mathrm{O}_2}^{\mathrm{RT}}(t)}{N_m^{\mathrm{test}}},
    \label{eq:sample_normalized_rate}
\end{equation}
where $N_m^{\mathrm{test}}$ is the test-sample quantity in Table~\ref{tab:rt_sample_quantities}: exposed area in $\mathrm{cm^2}$ for PTFE and PEEK, cable length in $\mathrm{m}$ for Kapton, and component count for PMT bases. The normalized rates are therefore expressed per $\mathrm{cm^2}$, per meter, or per component, respectively. This normalization is separate from the whole-sample load and from the physical surface-flux convention used in the diffusion model above.

Each material sample was measured inside the stainless-steel chamber at room temperature (approximately \SI{20}{\degreeCelsius}). The accumulation cycle was repeated automatically, with the chamber isolated for 15 s in each cycle, and measurements continued over several days. A common reference time of 72 h was chosen to match the practical pump-down timescale expected after prototype detector assembly while keeping the material-screening measurements experimentally feasible. To ensure reproducible initial conditions, all material samples were exposed to ambient air for more than one week before testing, allowing their surface-adsorbed gas content to approach an  equilibrium condition.

\begin{table}[t]
    \nolinenumbers
    \centering
    \captionsetup{justification=raggedright,singlelinecheck=false}
    \caption{
    Quantities of the samples used in the room-temperature outgassing
    measurements. The relevant scaling quantity is selected according
    to the geometry of each material or component.
    }
    \label{tab:rt_sample_quantities}
    \begin{tabular}{lcc}
        \hline
        Sample & Scaling quantity & Test-sample quantity \\
        \hline
        PTFE     & Exposed area & $5.01\times10^{2}~\mathrm{cm^2}$ \\
        PEEK     & Exposed area & $3.92\times10^{2}~\mathrm{cm^2}$ \\
        Kapton   & Cable length & $40~\mathrm{m}$ \\
        PMT base & Unit count   & 20 \\
        \hline
    \end{tabular}
\end{table}

\subsection{Room-temperature material outgassing measurements}
\label{sec:rt_measurements}

\begin{figure}[t]
    \nolinenumbers
    \centering
    \includegraphics[width=\columnwidth]{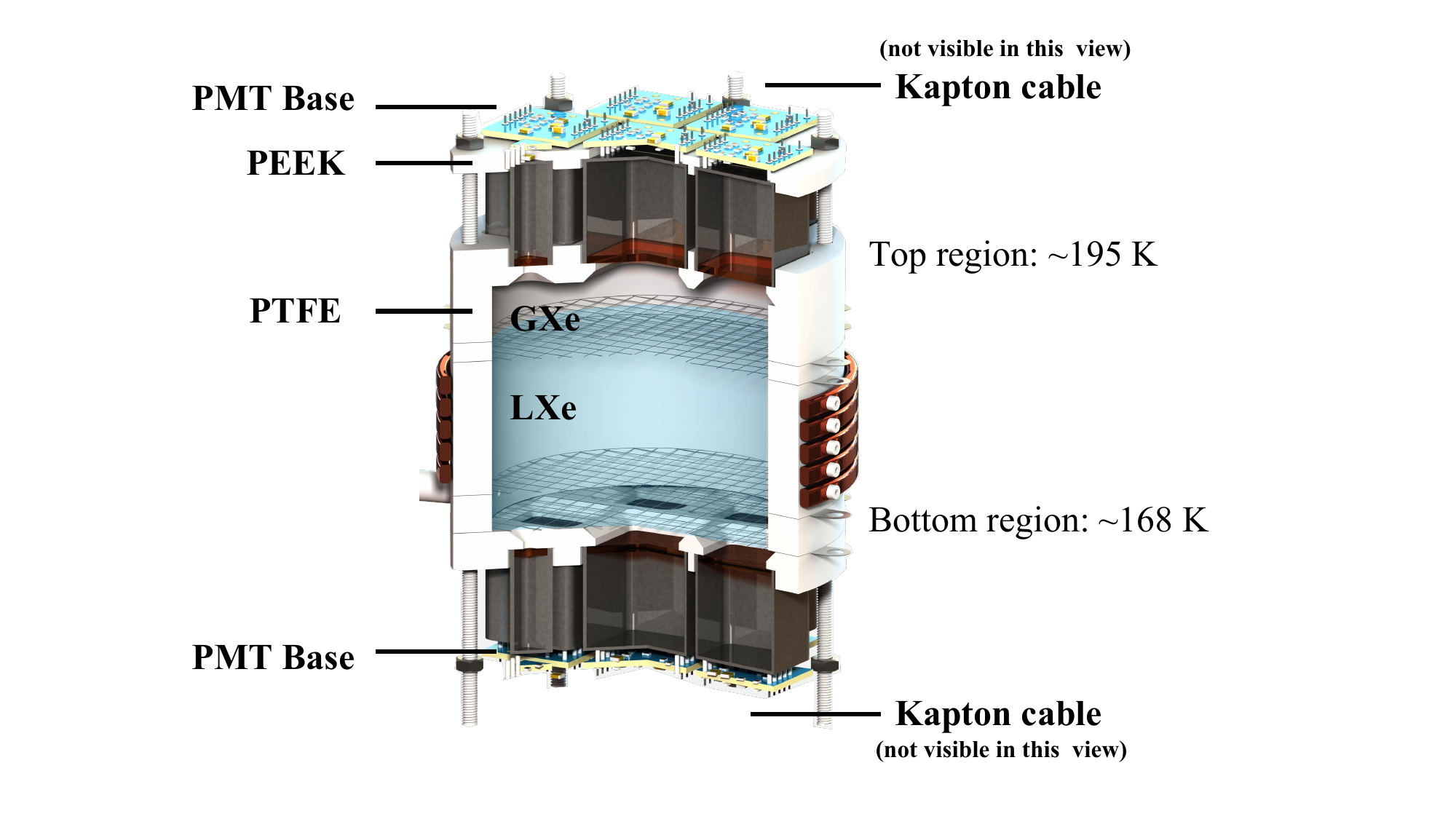}
    \captionsetup{justification=raggedright,singlelinecheck=false}
    \caption{
    Geometry of the RELICS prototype Run~7 TPC design, indicating the main non-metallic materials considered in the outgassing study. 
    PTFE is used as the inner reflector structure, while PMT bases are located at the top and bottom ends. 
    PEEK supports and Kapton cables are included in the detector assembly but are not fully visible in this cross-sectional view. 
    The representative temperatures of the upper and lower regions are approximately 195 K and 168 K, respectively~\cite{xie2026development}.
    }
    \label{fig:prototype_materials}
\end{figure}

Outgassing from non-metallic materials is expected to be a major contributor to the internal impurity load of the TPC because metals generally exhibit much lower room-temperature outgassing rates after standard vacuum cleaning and pumping procedures. In the RELICS prototype design, several non-metallic materials and components are used, including PTFE, PEEK, Kapton cables, and PMT bases, as indicated in the prototype Run~7 TPC geometry shown in Fig.~\ref{fig:prototype_materials} ~\cite{xie2026development}.
Since similar material and component choices are also planned for future RELICS detectors, their room-temperature oxygen outgassing rates were measured to provide input for estimating the impurity load arising from detector materials and guiding subsequent outgassing-suppression studies.

\begin{table*}[!t]
\nolinenumbers
\centering\small
\setlength{\tabcolsep}{4pt}
\caption{Room-temperature oxygen outgassing fits and rates at 72 h.}
\label{tab:rt_outgassing_fit}
\begin{tabular}{llcccc}
\hline
Sample & Range [h] & Model & $A$ & $b$ or $\tau$ [h] & $q_{72}$ \\
\hline
Empty chamber & 1.00--21.00 & $q_0+At^b$ & $8.74\times10^{3}\pm 1.8\times10^{2}$ & $-1.179\pm 0.014$ & -- \\
Empty chamber & 22.00--92.01 & $q_0+Ae^{-t/\tau}$ & $1.24\times10^{3}\pm 2.1\times10^{2}$ & $14.54\pm 1.4$ & $1042.9\pm 3.3$ \\
Kapton & 1.00--27.00 & $At^b$ & $1.01\times10^{6}\pm 6.6\times10^{4}$ & $-0.6618\pm 0.026$ & -- \\
Kapton & 28.00--72.00 & $Ae^{-t/\tau}$ & $1.87\times10^{5}\pm 2.3\times10^{3}$ & $47.01\pm 0.53$ & $4.05\times10^{4}\pm 2.7\times10^{2}$ \\
PEEK & 3.00--73.00 & $At^b$ & $9.31\times10^{5}\pm 1.8\times10^{4}$ & $-0.5098\pm 0.0054$ & $1.05\times10^{5}\pm 6.6\times10^{2}$ \\
PTFE & 1.00--81.00 & $At^b$ & $3.07\times10^{6}\pm 1.4\times10^{5}$ & $-0.639\pm 0.013$ & $1.99\times10^{5}\pm 3.2\times10^{3}$ \\
PMT base & 2.00--59.00 & $At^b$ & $7.62\times10^{5}\pm 2.7\times10^{4}$ & $-1.106\pm 0.01$ & -- \\
PMT base & 60.00--96.01 & $At^b$ & $4.59\times10^{4}\pm 7.5\times10^{3}$ & $-0.4163\pm 0.038$ & $7.74\times10^{3}\pm 45$ \\
\hline
\end{tabular}
\par\vspace{1mm}\begin{minipage}{0.98\textwidth}\footnotesize Time is in hours; rates use $\mathrm{ppb\,Pa\,L\,s^{-1}}$. In power laws, $t$ denotes its numerical value in hours. The empty-chamber plateau is $q_0=1034.2\pm 5.2$. Material rows use net rates. For piecewise fits, $q_{72}$ is reported once using the applicable late branch.\end{minipage}
\end{table*}

The measured time dependence was fitted in logarithmic space using the power-law and exponential models defined in Eqs.~\eqref{eq:outgassing_fit} and \eqref{eq:outgassing_fit_e}, with separate fits to early- and late-time regions where needed. The adopted models and ranges are listed in Table~\ref{tab:rt_outgassing_fit}; the quoted errors are conditional on these choices and the background treatment.

To determine the chamber background, the empty stainless-steel
chamber was first characterized using the same accumulation procedure. Since the
empty-chamber outgassing rate approaches a finite asymptotic value at long
times, an additional constant asymptotic background outgassing rate term was included in the
piecewise fit:
\begin{equation}
    q_{\mathrm{chamber}}(t)
    =
    q_{\mathrm{bg}} + f(t),
    \label{eq:empty_chamber_fit}
\end{equation}
where $q_{\mathrm{chamber}}(t)$ is the empty-chamber outgassing background, $f(t)$ represents the chamber outgassing component and $q_{bg}$ denotes the asymptotic background outgassing rate
\begin{equation}
    q_{\mathrm{bg}}
    =
    (1.0342\pm0.0052)\times10^{3}~\mathrm{ppb\,Pa\,L\,s^{-1}}.
    \label{eq:empty_chamber_floor}
\end{equation}

After the empty-chamber background was established, PTFE, PEEK, Kapton cables,
and PMT bases were measured individually under the same room-temperature
conditions. The quantities of the samples used in the accumulation measurements
are summarized in Table~\ref{tab:rt_sample_quantities}.

\begin{figure}[htbp]
    \nolinenumbers
    \centering
    \includegraphics[width=\columnwidth]
    {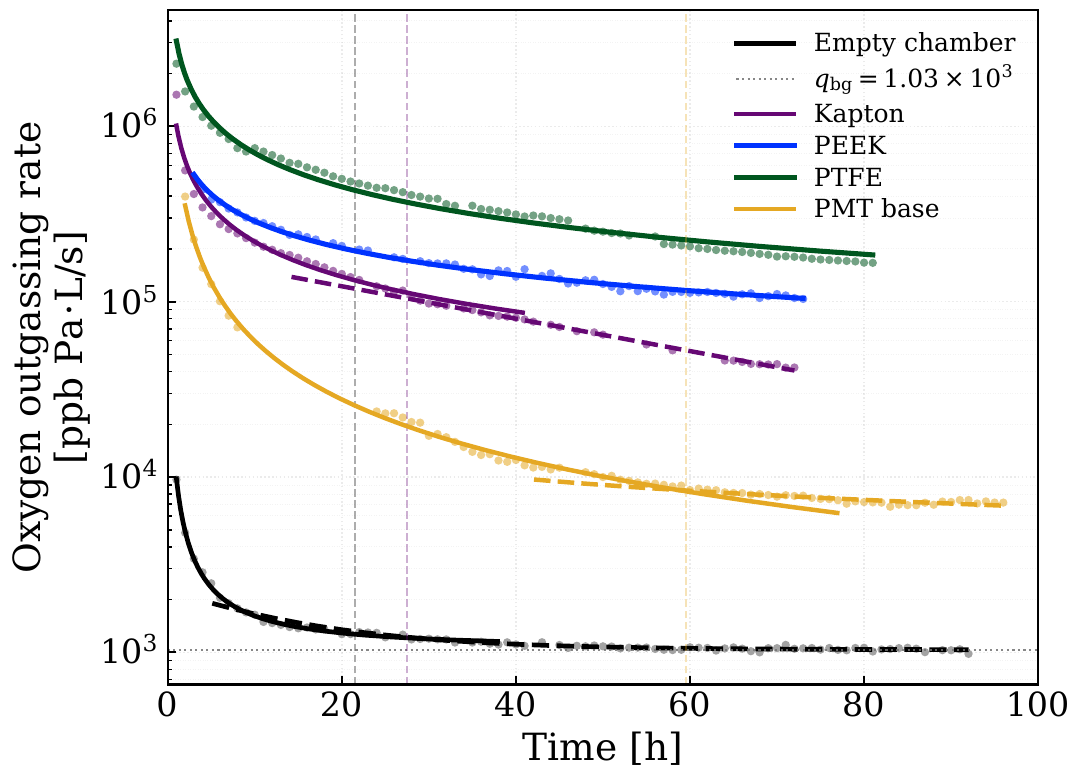}
    \captionsetup{justification=raggedright,singlelinecheck=false}
    \caption{
    Room-temperature oxygen outgassing rates measured for the empty
    chamber, Kapton cables, PEEK, PTFE, and PMT bases.
    Material curves show background-subtracted whole-sample rates.
    The markers represent the data, while the solid and dashed
    curves denote the fits in different time regions.
    The vertical dashed lines indicate the adopted transition times
    for datasets requiring piecewise fits, and the horizontal dotted
    line denotes the fitted asymptotic background outgassing rate $q_{bg}$.
    }
    \label{fig:outgassing_materials}
\end{figure}

The resulting room-temperature oxygen outgassing rates and the
corresponding fitted curves are shown in Fig.~\ref{fig:outgassing_materials}.
All material curves decrease with increasing pumping time. Background contributions were subtracted from the outgassing rates of all samples. PTFE and PEEK were adequately described by a single power law, whereas Kapton and the PMT bases required two time regions. The vertical dashed lines indicate the adopted boundaries. Table~\ref{tab:rt_outgassing_fit} summarizes the fitted parameters for all curves, including the empty-chamber background.

Several gaps are present in the measured time series because temporary
communication interruptions between the instruments and the
data-acquisition software resulted in missing data points. No interpolation was applied across these intervals, and only
the recorded valid data points were included in the fits. Small fluctuations are also visible in some datasets. They are likely related to environmental variations, such as changes in the ambient temperature, which can influence the chamber temperature, pressure readings, and material outgassing rates.

\subsection{Temperature Dependence of Material Outgassing}
\label{sec:cryo_measurements}

The material outgassing rate depends on both the elapsed pumping time
and the sample temperature. The temperature dependence must therefore be quantified before room-temperature rates can be used to estimate the material impurity load under the cryogenic operating conditions of RELICS. Under the assumption that these two effects can be approximately separated, it can be expressed schematically as 

\begin{equation}
    \Lambda(t,T)=F(t)G(T),
    \label{eq:time_temperature_dependence}
\end{equation}
where $F(t)$ describes the time-dependent depletion of the gas absorbed by the material, while $G(T)$ represents the temperature-dependent
transport, primarily governed by molecular diffusion within the material. The diffusion coefficient is commonly described by
an Arrhenius relation,
\begin{equation}
    D(T)
    =
    D_0
    \exp\left(-\frac{E_D}{RT}\right),
    \label{eq:arrhenius_diffusion}
\end{equation}
where $D_0$ is the pre-exponential factor, $E_D$ is the molar activation
energy for diffusion, $R$ is the molar gas constant, and $T$ is the
absolute temperature.

Under the ideal short-time diffusion approximation introduced in
Eq.~\eqref{eq:outgassing_short}, the outgassing rate is proportional to
$D^{1/2}$. Substituting Eq.~\eqref{eq:arrhenius_diffusion} gives
\begin{equation}
    \Lambda(t,T)
    \propto
    t^{-1/2}
    \exp\left(-\frac{E_D}{2RT}\right).
    \label{eq:ideal_time_temperature_relation}
\end{equation}
After accounting for the time-dependent contribution, the temperature dependence can be approximately
described by
\begin{equation}
    \Lambda(T)
    =
    \Lambda_0
    \exp\left(-\frac{E_{\mathrm{app}}}{RT}\right),
    \label{eq:arrhenius_outgassing}
\end{equation}
or equivalently,
\begin{equation}
    \ln \Lambda(T)
    =
    \ln \Lambda_0
    -
    \frac{E_{\mathrm{app}}}{R}\frac{1}{T},
    \label{eq:arrhenius_outgassing_linear}
\end{equation}
where $E_{\mathrm{app}}$ is the apparent activation energy obtained from the Arrhenius fitting
of the measured oxygen outgassing rate. In the ideal diffusion-limited
short-time approximation, this apparent activation energy is expected to satisfy $E_{\mathrm{app}}=E_D/2$. This relation is
used only as an approximate interpretation when comparing the measured
temperature dependence with reported diffusion activation energies.

The following analysis accounts for both the temperature dependence and the decay with pumping time, including a background contribution where it is non-negligible.

\subsubsection{Cryogenic Measurement System and Procedure}

\begin{figure}[t]
    \nolinenumbers
    \centering
    \includegraphics[width=\columnwidth]{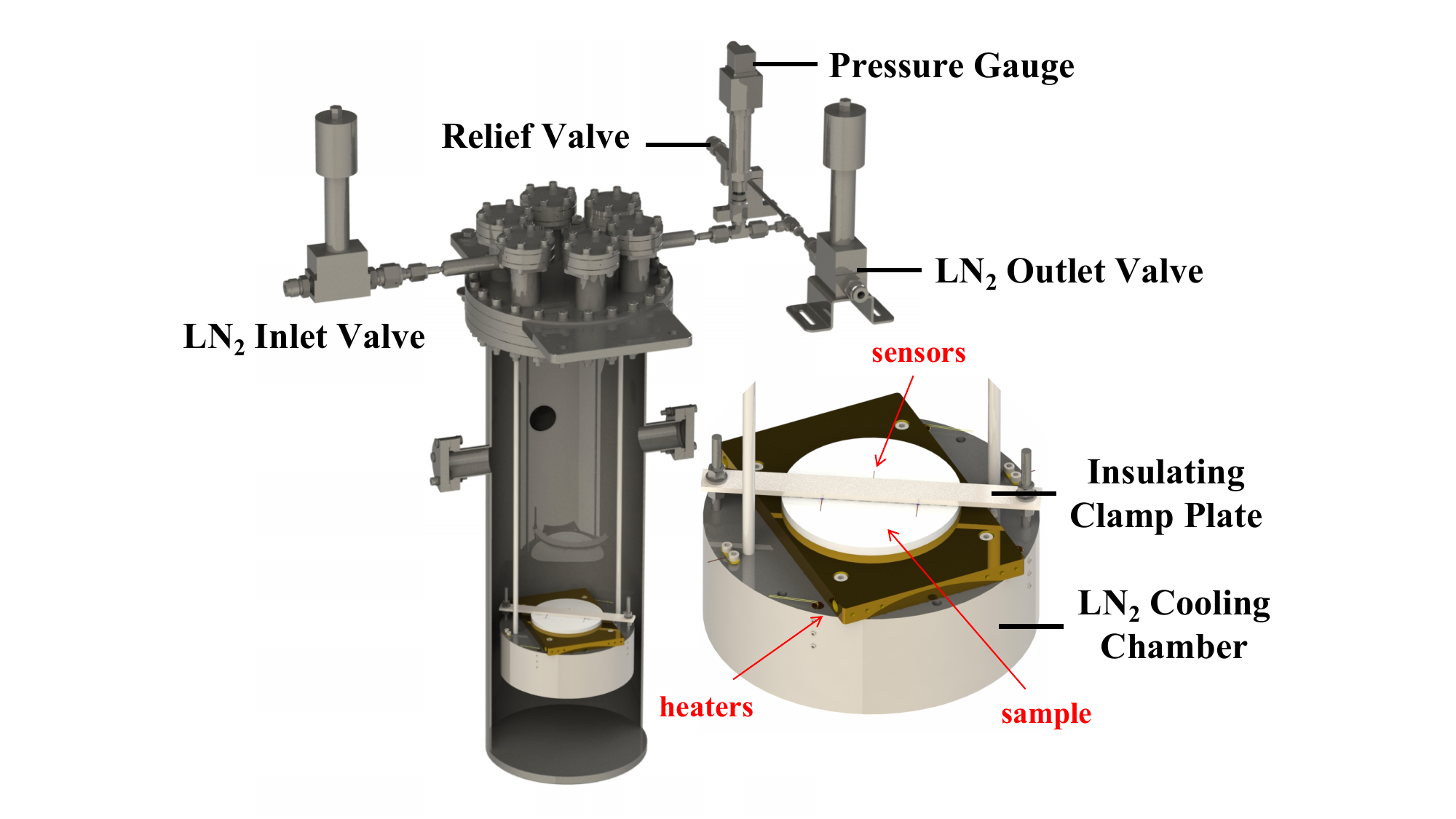}
    \captionsetup{justification=raggedright,singlelinecheck=false}
    \caption{
    Rendering of the cryogenic extension to the room-temperature
    outgassing measurement system. The enlarged view shows the LN$_2$
    cooling chamber, copper sample stage, insulating clamp plate,
    heaters, and temperature sensors. The LN$_2$ inlet and outlet lines
    are equipped with a pressure gauge and a relief valve.
    }
    \label{fig:cryo_system}
\end{figure}

To characterize oxygen outgassing under detector operating conditions,
a cryogenic module was integrated into the existing room-temperature measurement system,
as shown in Fig.~\ref{fig:cryo_system}. The module enables controlled
temperature-dependent outgassing measurements below room temperature while
retaining the original vacuum and gas-analysis system.

The central design objective of the cryogenic module is to provide a stable and controllable sample temperature under high-vacuum conditions. The LN$_2$ cooling chamber beneath the sample stage acts as a cryogenic cold reservoir, while heaters thermally coupled to the copper stage provide active thermal compensation against the cooling power. Multiple temperature sensors attached to the sample monitor its temperature and provide feedback to the PLC-based control system. Differences in thermal contact can cause differences between the sensor readings. By balancing the LN$_2$ cooling power and the applied heating power, the sample temperature can be adjusted and stabilized at the predefined set points. To ensure efficient thermal coupling between the cold stage and the sample, the sample is mechanically fixed on the copper stage using a low-thermal-conductivity clamp plate to minimize parasitic heat transfer, and low-outgassing vacuum-compatible cryogenic thermal grease is applied at the main contact interfaces.

\subsubsection{Outgassing Temperature Dependence Results}

Based on the room-temperature measurements, PTFE, PEEK, and Kapton were
selected for further cryogenic characterization because they represent
the principal polymeric materials contributing to the oxygen
outgassing in the TPC. Using the cryogenic measurement system shown in
Fig.~\ref{fig:cryo_system}, the samples were gradually cooled, and their
oxygen partial-pressure rise rates were measured during the cooling sequence. For convenience, these rates are used directly in $\mathrm{Pa\,s^{-1}}$; a common chamber-volume factor cancels in the temperature ratios and does not affect the Arrhenius slopes. Before Arrhenius fitting, the rates were corrected to the pumping age $t_0$ of the first point in each cooling sequence by multiplying by $F(t_0)/F(t)$, using the corresponding room-temperature decay fits and subtracting the background beforehand where applicable. This correction assumes approximately separable time and temperature dependences. The resulting temperature dependence
was analyzed using the Arrhenius representation introduced in
Eq.~\eqref{eq:arrhenius_outgassing_linear}. Depending on the observed
behavior, the data were fitted using either a single linear relation
or separate linear fits over different temperature ranges.

\begin{figure*}[!t]
    \nolinenumbers
    \centering

    \begin{subfigure}[t]{0.33\textwidth}
        \centering
        \includegraphics[width=\linewidth]
        {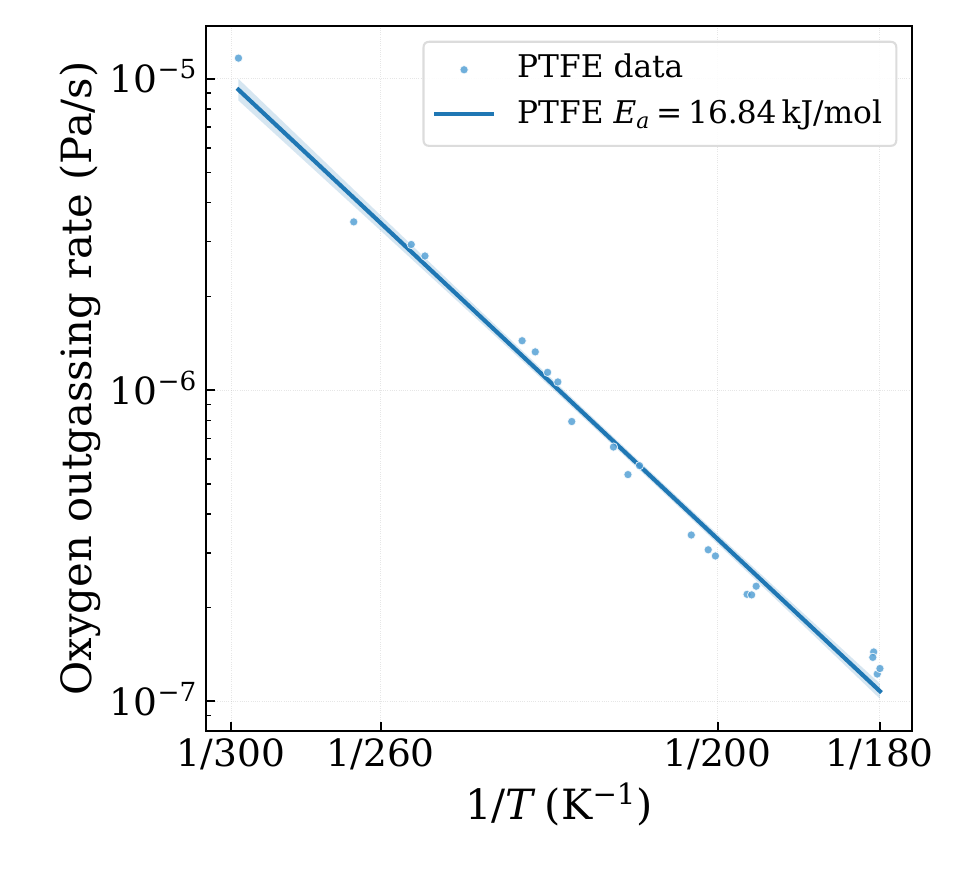}
        \caption{PTFE}
        \label{fig:ptfe_arrhenius}
    \end{subfigure}
    \hfill
    \begin{subfigure}[t]{0.33\textwidth}
        \centering
        \includegraphics[width=\linewidth]
        {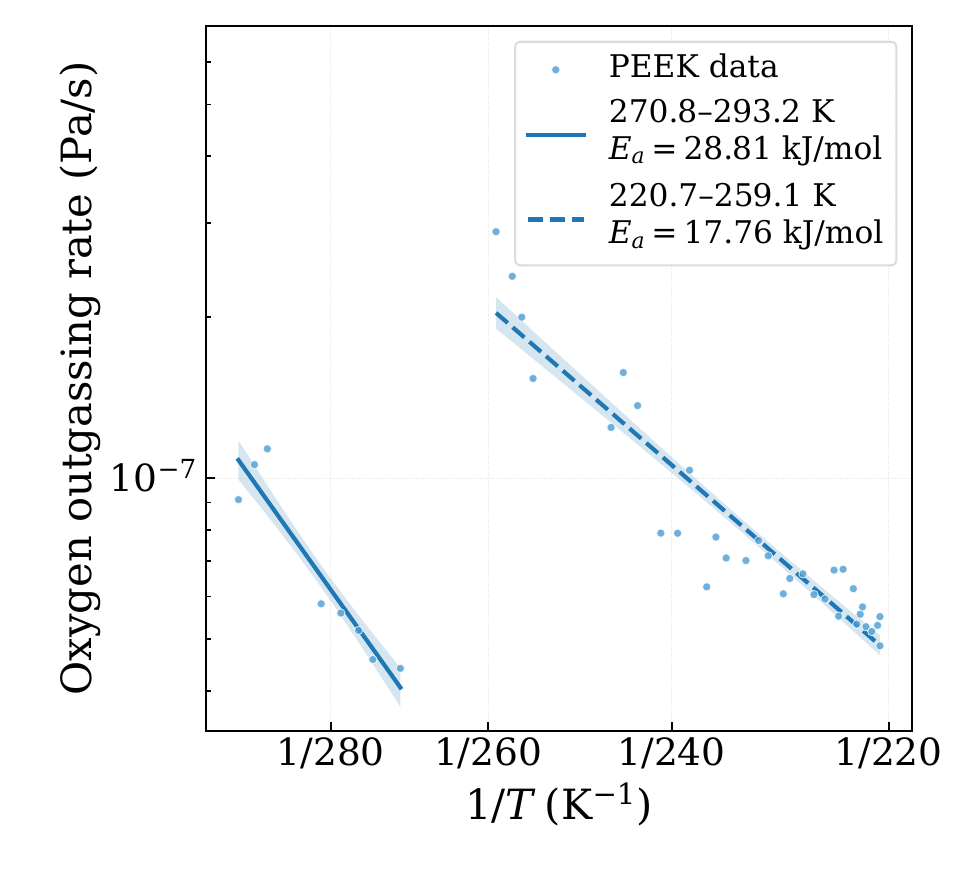}
        \caption{PEEK}
        \label{fig:peek_arrhenius}
    \end{subfigure}
    \hfill
    \begin{subfigure}[t]{0.33\textwidth}
        \centering
        \includegraphics[width=\linewidth]
        {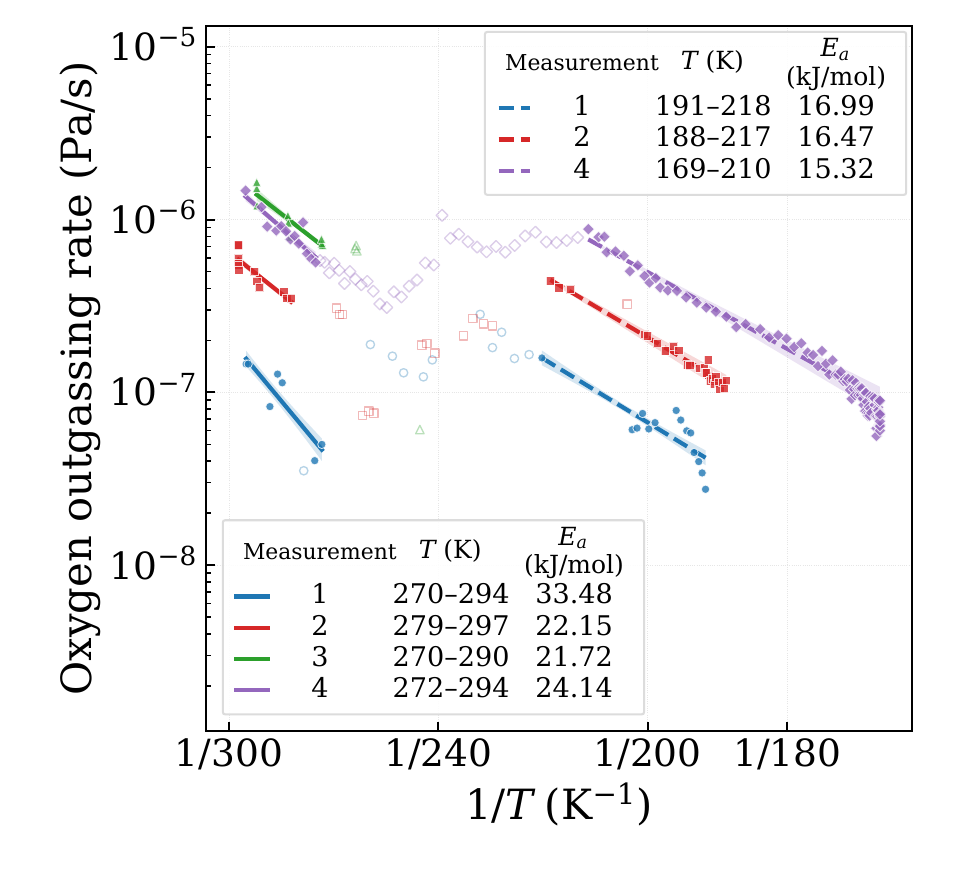}
        \caption{Kapton}
        \label{fig:kapton_arrhenius}
    \end{subfigure}
    \captionsetup{justification=raggedright,singlelinecheck=false}
    \caption{
    Arrhenius representations of the time-corrected oxygen
    partial-pressure rise rates of PTFE, PEEK, and Kapton. PTFE is described by a
    single linear relation, while PEEK and Kapton exhibit distinct
    high- and low-temperature regimes. Four repeated measurements on the same
    Kapton cable sample are shown.
    }
    \label{fig:material_arrhenius_results}
\end{figure*}

For PTFE, the oxygen outgassing rate decreases monotonically as the temperature is reduced over the measured range and is adequately described by a single linear fit in the Arrhenius representation, as shown in Fig.~\ref{fig:ptfe_arrhenius}. The fit gives an apparent activation energy of $E_{\mathrm{app}}^{\mathrm{PTFE}}=(16.84\pm0.45)~\mathrm{kJ\,mol^{-1}}$. Under the ideal short-time diffusion approximation, where the outgassing rate is proportional to $D^{1/2}$, this corresponds to an inferred diffusion activation energy of $E_D\simeq2E_{\mathrm{app}}=(33.67\pm0.90)~\mathrm{kJ\,mol^{-1}}$, which shows reasonable agreement with the oxygen diffusion activation energy of approximately $34.3~\mathrm{kJ\,mol^{-1}}$ reported for FEP Teflon~\cite{Koros1981Oxygen}. This consistency is compatible with a diffusion-controlled interpretation, although the comparison remains approximate because FEP Teflon and PTFE are different fluoropolymers and the reference value was obtained from direct diffusion-coefficient measurements over a different temperature range.

For PEEK, its Arrhenius behavior differs from that observed for PTFE. As shown in Fig.~\ref{fig:peek_arrhenius}, the data cannot be adequately described by a single linear relation over the full measured temperature range, and the regions above and below approximately $265~\mathrm{K}$ show different slopes. The data were therefore fitted separately in the high- and low-temperature regions, giving apparent activation energies of $E_{\mathrm{app}}^{\mathrm{PEEK,high}}=28.81~\mathrm{kJ\,mol^{-1}}$ for $T\geq265~\mathrm{K}$ and $E_{\mathrm{app}}^{\mathrm{PEEK,low}}=17.76~\mathrm{kJ\,mol^{-1}}$ for $T<265~\mathrm{K}$. The two-regime behavior suggests that the temperature dependence of oxygen outgassing from PEEK cannot be represented by a single effective activation energy over the entire measured range. Previous studies have reported multiple temperature-dependent relaxation processes in PEEK arising from different molecular motions and heterogeneous amorphous structures \cite{david1992peek,leonardi2013peek}. However, these studies do not establish the origin of the change observed here; the two fits are retained as an empirical description. The PEEK data shown in Fig.~\ref{fig:peek_arrhenius} extend down to approximately 221 K. Further cooling was limited by sample deformation, which reduced thermal contact with the copper stage.

\begin{table*}[!t]
\nolinenumbers
\centering\small
\setlength{\tabcolsep}{4pt}
\caption{Time-corrected Arrhenius fits and temperature factors relative to 293.15 K.}
\label{tab:temperature_dependence_summary}
\begin{tabular}{lcccc}
\hline
Material / measurement & Range [K] & $E_{\rm app}$ [kJ/mol] & $\eta(195\,\mathrm K)$ & $\eta(168\,\mathrm K)$ \\
\hline
PTFE & 180.0--297.7 & $16.84\pm 0.45$ & $0.0309\pm 0.0029$ & $0.00582\pm 0.00080$ \\
PEEK & 270.8--293.2 & $28.81\pm 4.1$ & $0.126\pm 0.018$ & $0.0217\pm 0.0053$ \\
PEEK & 220.7--259.1 & $17.76\pm 1.2$ & -- & -- \\
Kapton, measurement 1 & 270.1--294.0 & $33.48\pm 4.9$ & -- & -- \\
Kapton, measurement 2 & 279.3--296.6 & $22.15\pm 2.8$ & -- & -- \\
Kapton, measurement 3 & 269.9--290.4 & $21.72\pm 3.1$ & -- & -- \\
Kapton, measurement 4 & 271.9--294.1 & $24.14\pm 3.4$ & $0.298\pm 0.021$ & $0.0654\pm 0.012$ \\
Kapton, measurement 1 & 191.2--218.4 & $16.99\pm 2.2$ & -- & -- \\
Kapton, measurement 2 & 188.2--216.8 & $16.47\pm 1.2$ & -- & -- \\
Kapton, measurement 4 & 168.8--209.9 & $15.32\pm 1.4$ & -- & -- \\
\hline
\end{tabular}
\end{table*}

For Kapton, four repeated cooling measurements were performed on the same cable sample to assess reproducibility, since its wire geometry makes consistent thermal contact difficult. As shown in Fig.~\ref{fig:kapton_arrhenius}, the time-corrected rates are fitted separately in the high- and low-temperature regions. The time correction accounts for the rate decay within each cooling sequence but does not bring the four measurements to a same pumping age. Therefore, different pump-down durations before the measurements contribute to the differences in rate magnitude, while the fitted activation energies are broadly consistent, particularly at low temperatures. The apparent activation energies range from $21.72$ to $33.48~\mathrm{kJ\,mol^{-1}}$ at high temperatures and from $15.32$ to $16.99~\mathrm{kJ\,mol^{-1}}$ at low temperatures. Measurement~1 gives a higher high-temperature slope but is less constrained by its sparse sampling; measurement~3 does not provide a sufficiently constrained low-temperature branch. Under the ideal short-time diffusion approximation, the low-temperature values imply $E_D\simeq2E_{\mathrm{app}}=30.6$--$34.0~\mathrm{kJ\,mol^{-1}}$, close to the oxygen diffusion activation energy of $31.0~\mathrm{kJ\,mol^{-1}}$ reported for Kapton film~\cite{Koros1981Oxygen}. This comparison remains approximate because the reference measurements covered $137$--$237~{}^\circ\mathrm{C}$ and used film rather than cable insulation. Intermediate-temperature points without a clear Arrhenius trend were excluded from the fits. The downward deviation near the lowest temperature in measurement~4 may reflect sensitivity to background subtraction and residual temperature- or time-dependent effects not fully captured by the corrections.

Table~\ref{tab:temperature_dependence_summary} summarizes the fitted temperature ranges, apparent activation energies, and temperature factors $\eta(T)=\Lambda(T)/\Lambda(293.15~\mathrm{K})$ at the gas- and liquid-phase xenon temperatures of $195$ and $168~\mathrm{K}$. For piecewise fits, the factors use the applicable low-temperature branch and the high-temperature reference rate. The temperature factors from Kapton measurement~4 are adopted for the detector estimates without averaging the repeated measurements, because this measurement reaches the lowest temperature and its low-temperature slope agrees with measurement~2. These factors are used in the subsequent detector-level analysis.

\section{Vacuum Baking for Outgassing Reduction}
\label{sec:bakeout_mitigation}

The purity model indicates that reducing the material outgassing load
can effectively improve the electron lifetime expected for the RELICS
detector~\cite{yu2026projection}. Considering the room-temperature
outgassing measurements, the temperature-dependent cryogenic results,
and the material inventory in the detector, PTFE is expected to be
an important contributor to the oxygen outgassing source term. Therefore,
reducing the oxygen outgassing rate of PTFE before detector assembly is
an important step toward improving the achievable purity performance.

Vacuum baking was therefore selected as a practical treatment to reduce
the oxygen outgassing rate of PTFE before detector assembly. Bake-out is
widely used in vacuum and contamination-control applications to
accelerate gas release by heating materials, often under vacuum, thereby
reducing the subsequent outgassing load~\cite{Grinham2017Outgassing,
NASA_Bakeout}. By heating PTFE under vacuum, gas retained inside the
polymer can be released more rapidly, reducing the residual gas
inventory responsible for subsequent outgassing. In addition to the
baking temperature and duration, the air-exposure time after baking must
also be considered, because PTFE components are temporarily exposed to
air between removal from the vacuum oven and installation into the
detector. Gas release from polymers after exposure to humid air has also been studied in vacuum systems~\cite{Sammartano2020Polymers}.
Therefore, the treatment study focuses on two practical issues: the
baking conditions needed to reduce the PTFE oxygen outgassing rate and
the extent to which this reduction can be retained after a representative
post-bake air-exposure period.

PTFE samples were baked for 100 and 283 h at \SI{473}{K} in a vacuum oven connected to a
mechanical pump, with the oven pressure maintained at the Pa level
during baking. A shorter treatment at \SI{393}{K} for 3 h 40 min was included for comparison. After baking, the PTFE sample was promptly installed in the outgassing measurement chamber to minimize exposure to ambient air before measurement. To evaluate the practical
effect of air exposure during TPC assembly, an additional PTFE
sample baked for 322 h at \SI{473}{K} was exposed to ambient air for \(48~\mathrm{h}\), corresponding
to the expected assembly time scale, and was then measured using the
same outgassing setup.

\begin{table*}[!t]
\nolinenumbers
\centering\small
\setlength{\tabcolsep}{3pt}
\caption{PTFE treatment fits, rates at 72 h, and ratios to untreated PTFE.}
\label{tab:ptfe_bake_fit}
\begin{tabular}{llccccc}
\hline
Treatment & Range [h] & Model & $A$ & $b$ or $\tau$ [h] & $q_{72}$ & $B_{72}$ \\
\hline
PTFE, untreated & 1.00--81.00 & $At^b$ & $3.07\times10^{6} \pm 1.4\times10^{5}$ & $-0.639 \pm 0.013$ & $1.99\times10^{5} \pm 3.2\times10^{3}$ & $1$ \\
PTFE, 3 h 40 min & 1.00--98.01 & $At^b$ & $7.67\times10^{5} \pm 3.0\times10^{4}$ & $-0.4416 \pm 0.01$ & $1.16\times10^{5} \pm 1.4\times10^{3}$ & $0.582 \pm 0.012$ \\
PTFE, 100 h & 1.00--149.01 & $At^b$ & $7.22\times10^{4} \pm 2.9\times10^{3}$ & $-0.9213 \pm 0.01$ & $1.40\times10^{3} \pm 16$ & $0.00703 \pm 1.4\times10^{-4}$ \\
PTFE, 283 h & 1.00--96.01 & $At^b$ & $1.50\times10^{4} \pm 5.3\times10^{2}$ & $-0.6112 \pm 0.0095$ & $1.10\times10^{3} \pm 12$ & $0.0055 \pm 1.1\times10^{-4}$ \\
\begin{tabular}[c]{@{}l@{}}PTFE, 322 h\\+48 h air\end{tabular} & 8.00--48.00 & $At^b$ & $4.30\times10^{6} \pm 3.0\times10^{5}$ & $-1.094 \pm 0.021$ & $4.00\times10^{4} \pm 1.0\times10^{3}$ & $0.200 \pm 0.006$ \\
\hline
\end{tabular}
\par\vspace{1mm}\begin{minipage}{0.98\textwidth}\footnotesize Time is in hours; rates use $\mathrm{ppb\,Pa\,L\,s^{-1}}$. In power laws, $t$ denotes its numerical value in hours. $B_{72}=q_{\mathrm{treated}}(72)/q_{\mathrm{untreated}}(72)$; this is time-specific. The empty-chamber fits are given in Table~\ref{tab:rt_outgassing_fit}.\end{minipage}
\end{table*}

\begin{figure}[t]
    \nolinenumbers
    \raggedright
    \includegraphics[width=\columnwidth]
    {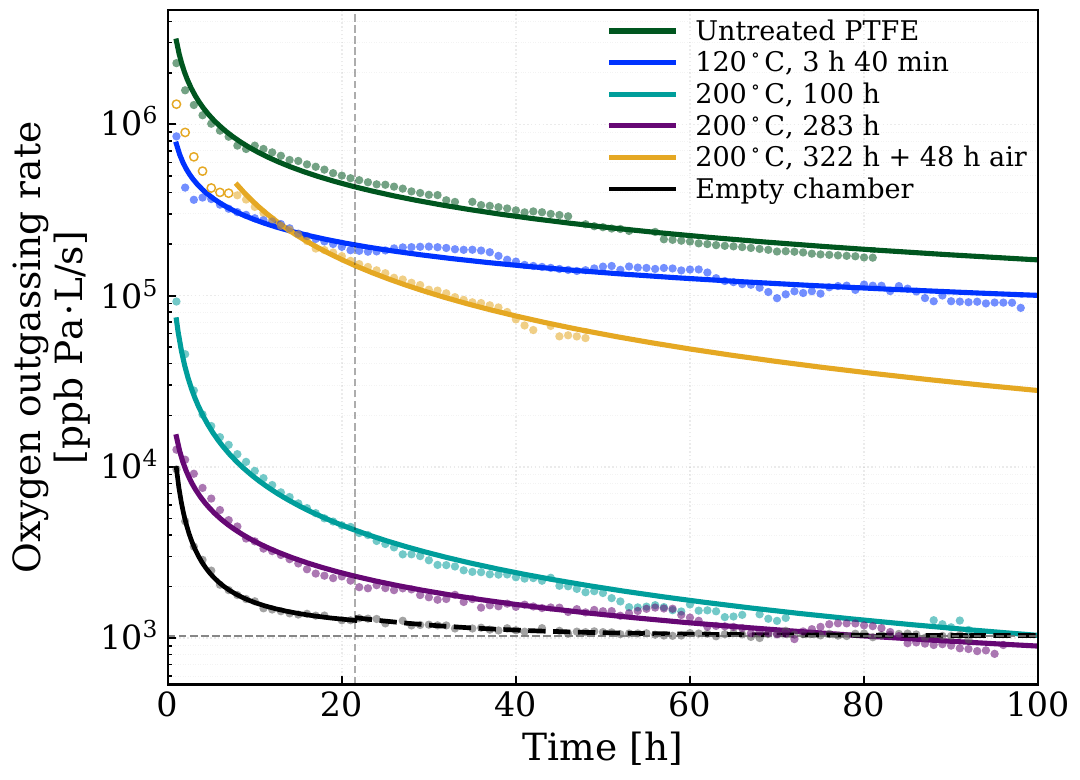}
    \captionsetup{
        justification=raggedright,
        singlelinecheck=false
    } 
    \caption{
        Oxygen outgassing rates of PTFE under different baking and
        post-bake exposure conditions. 
        Material curves show the background-subtracted contributions.
    }
    \label{fig:ptfe_bake_duration}
\end{figure}

The outgassing curves obtained under different baking and exposure
conditions are shown in Fig.~\ref{fig:ptfe_bake_duration}, and the
corresponding fit results are summarized in
Table~\ref{tab:ptfe_bake_fit}. The material curves were fitted with functions appropriate to their time dependence and are shown after background subtraction. The results show that prolonged vacuum baking at
\(473~\mathrm{K}\) substantially suppresses oxygen outgassing from
PTFE. Although the overall outgassing level decreases with increasing
baking duration, the directly measured baked PTFE curves retain a
power-law decay, consistent with a reduced gas inventory. After \(283~\mathrm{h}\) of baking, the PTFE outgassing rate
at long pumping times is reduced by more than two orders of magnitude
relative to the unbaked sample and reaches a level comparable to the
stainless-steel chamber background. Based on this result, a baking
duration of about \(300~\mathrm{h}\) was adopted for subsequent detector
preparation.

The \(48~\mathrm{h}\) air-exposure test shows the practical impact of
assembly handling after baking. As shown in
Fig.~\ref{fig:ptfe_bake_duration}, the air-exposed sample has a higher
oxygen outgassing rate than the promptly measured long-baked samples,
consistent with renewed gas uptake during air exposure, but remains below
the unbaked sample. Exposure during assembly should nevertheless be minimized. The fitted
48 h exposure result is used in the following detector-level estimate.

\section{Detector-Level Outgassing Estimation and Projection}
\label{sec:detector_level_prediction}

The material-level measurements described above are converted into
detector-level oxygen-equivalent source terms in two steps. First, the
oxygen-equivalent gas load introduced into detector volume $i$ is calculated by
scaling the room-temperature specific outgassing rate of each material by
its quantity in the detector, its cryogenic suppression factor, and, when
applicable, its time-dependent treatment factor:
\begin{equation}
    Q_i^{\mathrm{O}_2}(t,T_i)
    =
    \sum_m
    N_{i,m}\,
    \Lambda_{m,\mathrm{O}_2}^{\mathrm{RT}}(t)\,
    \eta_m(T_i)\,
    B_{m}(t),
    \label{eq:detector_oxygen_load}
\end{equation}
where $N_{i,m}$ denotes the quantity of material or component $m$ exposed to volume $i$, and $\Lambda_{m,\mathrm{O}_2}^{\mathrm{RT}}(t)$ is its room-temperature oxygen-equivalent outgassing rate per unit of the corresponding quantity after a pump-down time $t$. The dimensionless factor $\eta_m(T_i)$ accounts for the suppression of outgassing at the detector temperature $T_i$, and $B_{m}(t)$ is the treated-to-untreated ratio at the same pumping time, where applicable. 

For PTFE and PEEK,
$N_{i,m}$ is an exposed surface area in $\mathrm{cm^2}$, whereas for Kapton it is a cable
length in meters. Component-level measurements, such as those of PMT bases, are
normalized by unit count. The units used for $N_{i,m}$ match those
used to normalize $\Lambda_{m,\mathrm{O}_2}^{\mathrm{RT}}$.

The temperature-reduction factor is defined as
\begin{equation}
    \eta_m(T_i)
    =
    \frac{\Lambda_{m,\mathrm{O}_2}(T_i)}
         {\Lambda_{m,\mathrm{O}_2}(T_{\mathrm{ref}})},
    \qquad
    T_{\mathrm{ref}} = 293.15~\mathrm{K},
    \label{eq:temperature_reduction_factor}
\end{equation}
and is obtained from the measured temperature dependences summarized in
Table~\ref{tab:temperature_dependence_summary}. Applying the same temperature factor to treated PTFE assumes that treatment does not change its temperature dependence. The baking-reduction factor is defined
relative to the corresponding unbaked material:
\begin{equation}
    B_{m}(t)
    =
    \frac{
        \Lambda_{m,\mathrm{O}_2}^{\mathrm{baked,RT}}(t)
    }{
        \Lambda_{m,\mathrm{O}_2}^{\mathrm{unbaked,RT}}(t)
    }.
    \label{eq:baking_reduction_factor}
\end{equation}
For untreated materials, $B_m=1$. When a single
reference pump-down time is used for the detector-level comparison, the
baking factor is evaluated at the same time, namely
$t_{\mathrm{ref}}=72~\mathrm{h}$.

After scaling the material-level outgassing rates by the detector material inventory and applying the relevant correction factors, the resulting detector-level oxygen-equivalent gas load,
$Q_i^{\mathrm{O}_2}$, follows the same $10^9$-scaled reporting convention as the sample loads and has units of
\si{\ppb\pascal\liter\per\second}. It is converted to the source
term used in the RELICS purity-evolution model according to
\begin{equation}
    f_i^{\mathrm{mat}}(t,T_i)
    =
    \frac{3.6}{R T_{\mathrm{gas}}}\,
    Q_i^{\mathrm{O}_2}(t,T_i),
    \label{eq:detector_source_term}
\end{equation}
where $R$ is the universal gas constant expressed in SI units. We adopt the same nominal room temperature, $T_{\mathrm{gas}}=T_{\mathrm{ref}}=293.15~\mathrm{K}$, for the gas-load conversion and the reference of $\eta_m$. The detector material temperature $T_i$ enters through $\eta_m(T_i)$. The factor $3.6$ accounts for
conversion from liters to cubic meters and from seconds to hours.
The resulting $f_i^{\mathrm{mat}}$ is expressed in
\si{\ppb\mole\per\hour}, consistent with the definition of the
outgassing source terms in the RELICS purity-evolution
model.

\subsection{Comparison with Run~7 and Run~9}
\label{sec:run7_run9_comparison}

\begin{figure}[htbp]
    \nolinenumbers
    \centering
    \begin{minipage}{0.75\linewidth}
        \centering
        \textbf{(a) Run~7: Diving-bell configuration}\par
        \smallskip
        \includegraphics[width=\linewidth]{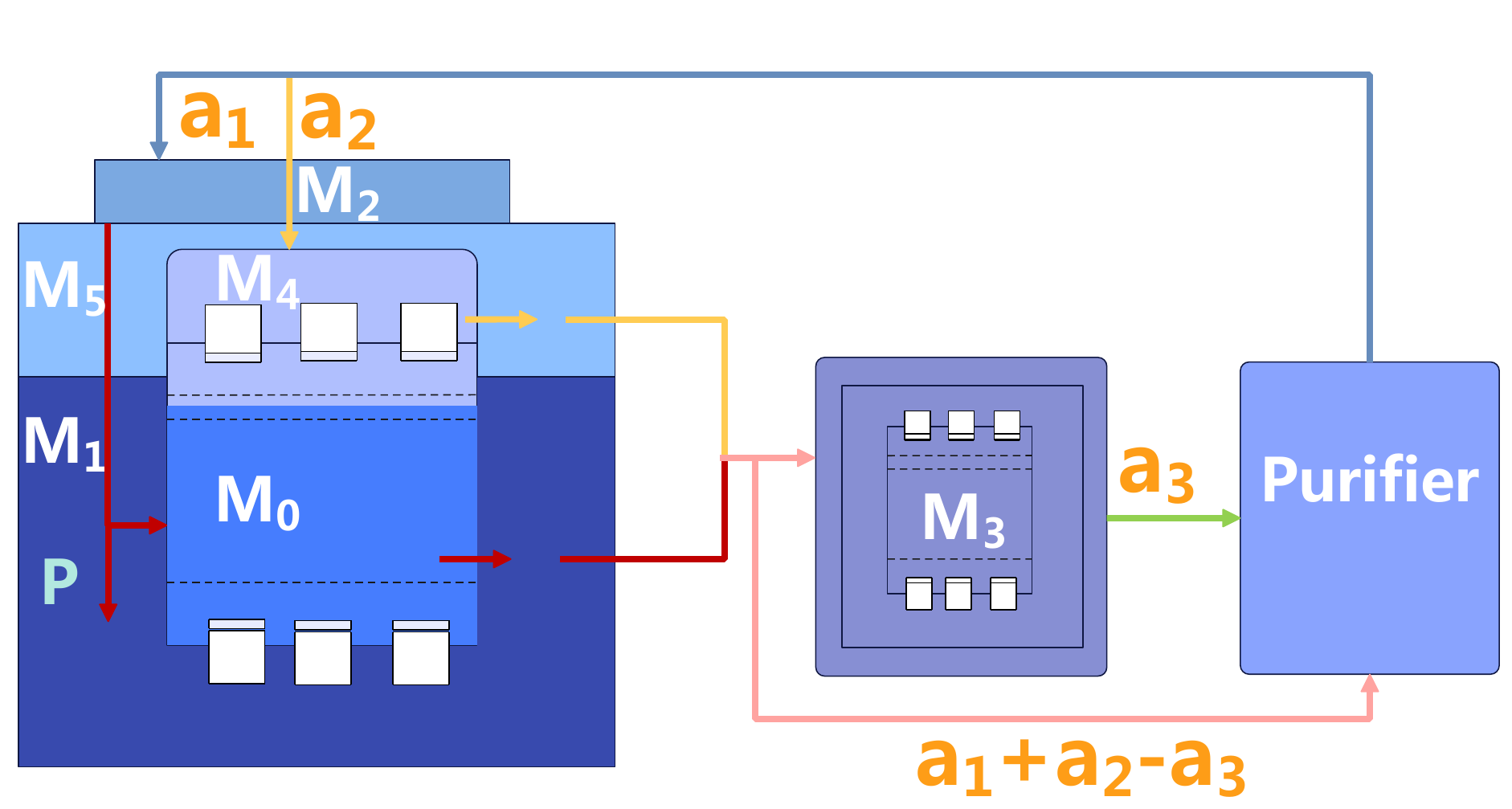}
    \end{minipage}

    \vspace{0.5em}

    \begin{minipage}{0.75\linewidth}
        \centering
        \textbf{(b) Run~9: Overflow-chamber configuration}\par
        \smallskip
        \includegraphics[width=\linewidth]{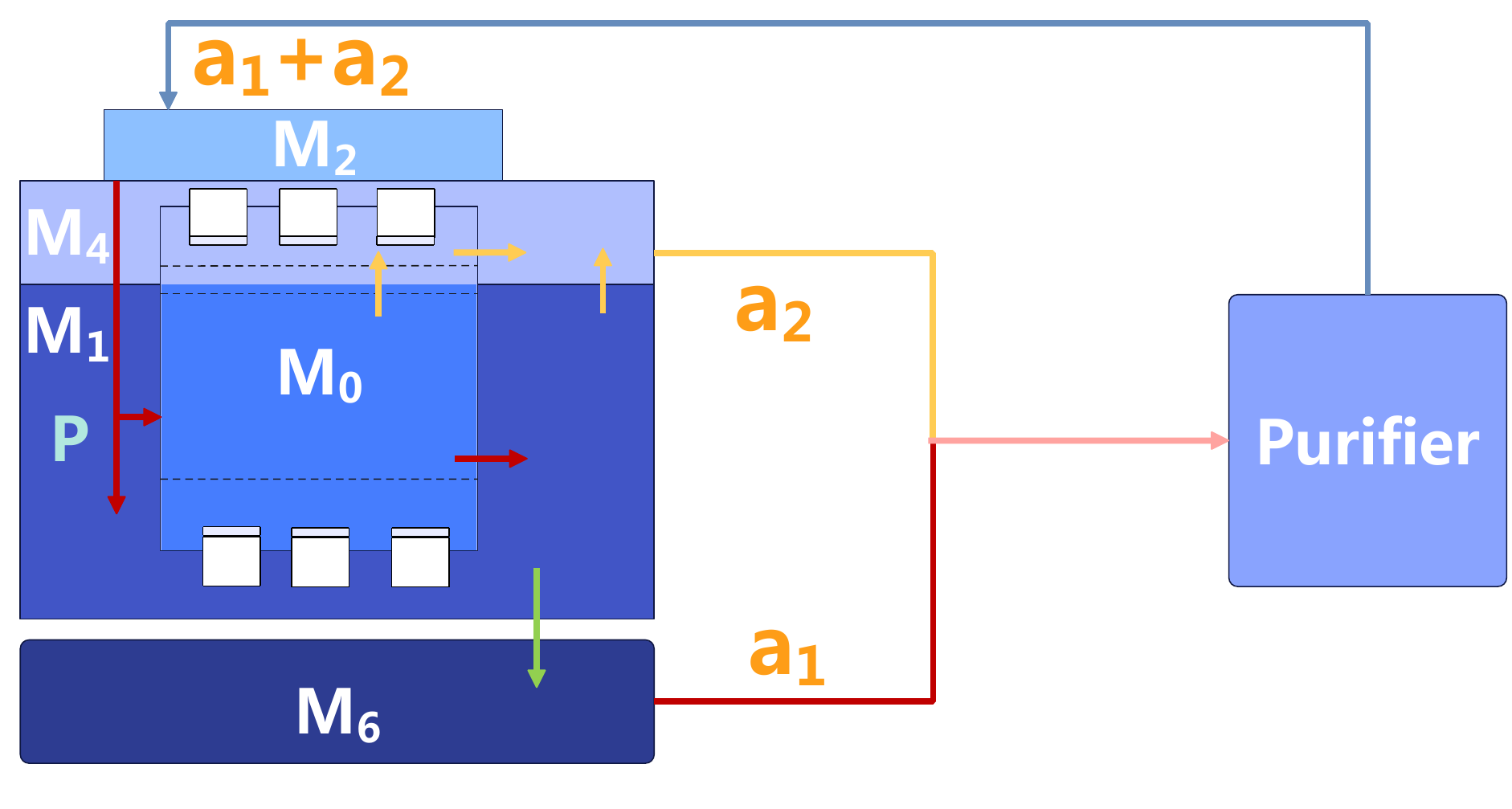}
    \end{minipage}

    \caption{
        Schematics of the Run~7 and Run~9 prototype configurations.
        Reproduced from Figs.~8 and~9 of Ref.~\cite{yu2026projection}.
    }
    \label{fig:prototype_control_volumes}
\end{figure}

To cross-check the material-based estimates, we compare the source terms at $t=72~\mathrm{h}$ with those inferred from the Run~7 and Run~9 purity-evolution fits~\cite{yu2026projection}. Run~7 used PTFE as the main structural polymer, whereas Run~9 used PEEK. Table~\ref{tab:prototype_inventory} summarizes their material inventories, temperatures, and source terms, calculated using Eqs.~\eqref{eq:detector_oxygen_load} and~\eqref{eq:detector_source_term}. The quoted uncertainties include the room-temperature fits and temperature factors. Figure~\ref{fig:prototype_control_volumes} shows the control volumes; their labels follow the cited prototype definitions and do not necessarily denote the same physical region in both runs.

\begin{table*}[!t]
\nolinenumbers
\centering\small
\setlength{\tabcolsep}{2.7pt}
\caption{Prototype material inventories and source terms in Run~7 and Run~9.}
\label{tab:prototype_inventory}
\label{tab:prototype_source_comparison}
\begin{tabular}{cccccccccc}
\hline
\multirow{2}{*}{Detector} & \multirow{2}{*}{Region} & \multirow{2}{*}{\shortstack{Temperature\\{[K]}}} & \multirow{2}{*}{Polymer} & \multirow{2}{*}{\shortstack{Area\\{[cm$^2$]}}} & \multirow{2}{*}{\shortstack{Kapton length\\{[m]}}} & \multicolumn{4}{c}{Source term [ppb mol h$^{-1}$]} \\
\cline{7-10}
 & & & & & & Polymer & Kapton & Total & Purification fit \\
\hline
Run~7 & M0 & 168 & PTFE & 557 & 0 & $1.91\pm 0.26$ & -- & $1.91\pm 0.26$ & $1.9\pm 0.2$ \\
Run~7 & M1 & 168 & PTFE & 3342 & 2 & $11.4\pm 1.6$ & $0.196\pm 0.035$ & $11.6\pm 1.6$ & $11.4\pm 1.7$ \\
Run~7 & M4 & 195 & PTFE & 95 & 0 & $1.73\pm 0.16$ & -- & $1.73\pm 0.16$ & $7\pm 7.5$ \\
Run~7 & M5 & 195 & PTFE & 230 & 18 & $4.18\pm 0.39$ & $8.04\pm 0.56$ & $12.2\pm 0.69$ & $18.4\pm 12$ \\
Run~9 & M0 & 168 & PEEK & 867 & 0 & $7.45\pm 1.8$ & -- & $7.45\pm 1.8$ & $3.2\pm 0.9$ \\
Run~9 & M1 & 168 & PEEK & 1142 & 2 & $9.81\pm 2.4$ & $0.196\pm 0.035$ & $10\pm 2.4$ & $3.4\pm 1.4$ \\
Run~9 & M4 & 195 & PEEK & 414 & 23 & $20.7\pm 3$ & $10.3\pm 0.72$ & $30.9\pm 3.1$ & $8.9\pm 1.5$ \\
\hline
\end{tabular}
\end{table*}

Table~\ref{tab:prototype_source_comparison} shows generally compatible source terms for Run~7 within the quoted uncertainties, while the material estimates are systematically higher for Run~9. The PTFE measurements extend down to 180 K, close to the operating temperatures of 168--195 K, providing a more direct constraint on the Run~7 estimate. In contrast, the PEEK measurements reach only 221 K, so the Run~9 estimate requires extrapolation over a wider temperature interval. The PEEK-based estimate is also sensitive to the room-temperature background treatment and to differences between vacuum testing and detector operation. These effects are possible contributors, but their individual roles are not established by the present comparison.

\begin{table*}[!t]
\nolinenumbers
\centering\small
\setlength{\tabcolsep}{4pt}
\caption{Material inventories and predicted oxygen-equivalent source terms for RELICS-10 and RELICS-50. The PTFE source terms include the vacuum baking at 473 K for 322 h followed by 48 h of air exposure.}
\label{tab:future_inventory}
\begin{tabular}{ccccccc}
\hline
Detector & Region & Temperature [K] & Polymer & Area [cm$^2$] & Kapton length [m] & Total source term [ppb mol h$^{-1}$] \\
\hline
RELICS-10 & M0 & 168 & PTFE & 1323 & 0 & $0.908\pm 0.13$ \\
RELICS-10 & M1 & 168 & PTFE & 1323 & 12 & $2.08\pm 0.25$ \\
RELICS-10 & M4 & 195 & PTFE & 755 & 80 & $38.5\pm 2.5$ \\
RELICS-50 & M0 & 168 & PTFE & 4709 & 0 & $3.23\pm 0.45$ \\
RELICS-50 & M1 & 168 & PTFE & 4709 & 64 & $9.49\pm 1.2$ \\
RELICS-50 & M4 & 195 & PTFE & 2559 & 288 & $138\pm 9$ \\
\hline
\end{tabular}
\end{table*}

\subsection{Projection for Future RELICS Detectors}
\label{sec:future_relics_projection}

The material-based source-term estimation is also applied to the planned RELICS-10 and RELICS-50 detectors. Both designs adopt the overflow-chamber configuration shown in Fig.~\ref{fig:prototype_control_volumes}(b), with the same definitions of the regions $M_n$. Table~\ref{tab:future_inventory} lists their inventories and source terms at 72 h, calculated using Eqs.~\eqref{eq:detector_oxygen_load} and~\eqref{eq:detector_source_term} with the treatment factor $B_{72}$ in Table~\ref{tab:ptfe_bake_fit} for PTFE baked at \SI{473}{K} for 322 h and subsequently exposed to air for 48 h. These estimated source terms can be used as inputs to the purification model of Ref.~\cite{yu2026projection} to predict the time evolution of xenon purity in RELICS.

\FloatBarrier
\section{Summary}
\label{sec:sum}

The RELICS experiment requires high liquid-xenon purity to detect low-energy nuclear recoils from reactor-antineutrino CE$\nu$NS. Quantifying and suppressing material outgassing is therefore essential to limiting electronegative impurities. In this work, we measured the oxygen-equivalent outgassing rates of detector materials, characterized their temperature dependence, and investigated vacuum baking as a means of mitigating material-related impurity sources in RELICS. Prolonged vacuum baking substantially reduced PTFE outgassing. Subsequent air exposure weakened this benefit, highlighting the need to minimize exposure during detector assembly. The room-temperature and low-temperature measurements were used to estimate detector-level source terms for the Run~7 and Run~9 prototypes. Comparison with the source terms inferred from the purity-evolution model fits showed general agreement for Run~7, while the material-based estimates were higher for Run~9. The same method was applied to the planned RELICS-10 and RELICS-50 detectors using the fitted post-baking, air-exposed PTFE rates. These estimates provide material-source inputs for future purity projections under the studied preparation conditions.

\section*{Acknowledgments}
We are grateful to Prof. Xiaopeng Zhou from Beihang University for providing the $^{37}$Ar and $^{83\text{m}}\text{Kr}$ sources used for the calibration of the RELICS prototype. We also thank Jiayi Li of the Department of Astronomy at Tsinghua University for assistance with the hardware design.

RELICS is supported by grants from National Key R\&D program from the Ministry of Science and Technology of China (No. 2021YFA1601600), Natural Science Foundation of China (Nos. 12275267, 12521007, 12405129, 12375095, 12250011, 12305121, W2543014), Ministry of Education of China (No. SRICSPYF-ZY2025028), Beijing Natural Science Foundation (Nos. QY23088, QY25008), Guangzhou Municipal Science and Technology Project (No. 2025A04J5409), Zhejiang Provincial Natural Science Foundation of China (No. LQKWL26A0501), Pengcheng Peacock Program Supporting Research Funding (Class C, No. KQ002778), and the CUHK-Shenzhen University Development Fund (No. UDF01003491).



\begin{thebibliography}{99}

\bibitem{Freedman1974}
D. Z. Freedman, Coherent neutrino nucleus scattering as a probe of the weak neutral current, \emph{Phys. Rev. D} \textbf{9}, 1389--1392 (1974). https://doi.org/10.1103/PhysRevD.9.1389

\bibitem{canas2018future}
BC Ca{\~n}as, et al., Future perspectives for a weak mixing angle measurement in coherent elastic neutrino nucleus scattering experiments, \emph{Physics Letters B} \textbf{784}, 159--162 (2018).

\bibitem{patton2012neutrino}
Kelly Patton, et al., Neutrino-nucleus coherent scattering as a probe of neutron density distributions, \emph{Physical Review C—Nuclear Physics} \textbf{86}, 024612 (2012).


\bibitem{beacom2010diffuse}
John F Beacom, The diffuse supernova neutrino background, \emph{Annual Review of Nuclear and Particle Science} \textbf{60}, 439--462 (2010).

\bibitem{Drukier1984}
A. Drukier, L. Stodolsky, Principles and applications of a neutral-current detector for neutrino physics and astronomy, \emph{Phys. Rev. D} \textbf{30}, 2295--2309 (1984). https://doi.org/10.1103/PhysRevD.30.2295

\bibitem{coloma2017curtailing}
Pilar Coloma, et al., Curtailing the dark side in non-standard neutrino interactions, \emph{Journal of High Energy Physics} \textbf{2017}, 116 (2017).



\bibitem{COHERENT_CsI}
D. Akimov, et al. (COHERENT Collaboration), Observation of coherent elastic neutrino-nucleus scattering, \emph{Science} \textbf{357}, 1123--1126 (2017). https://doi.org/10.1126/science.aao0990

\bibitem{COHERENT_LAr}
D. Akimov, et al. (COHERENT Collaboration), First Measurement of Coherent Elastic Neutrino-Nucleus Scattering on Argon, \emph{Phys. Rev. Lett.} \textbf{126}, 012002 (2021). https://doi.org/10.1103/PhysRevLett.126.012002

\bibitem{CONUSplus}
N. Ackermann, et al. (CONUS+ Collaboration), Direct observation of coherent elastic antineutrino--nucleus scattering, \emph{Nature} \textbf{643}, 1229--1233 (2025). https://doi.org/10.1038/s41586-025-09322-2

\bibitem{XENONnT_CEvNS}
E. Aprile, et al. (XENON Collaboration), First Indication of Solar $^{8}\mathrm{B}$ Neutrinos via Coherent Elastic Neutrino-Nucleus Scattering with XENONnT, \emph{Phys. Rev. Lett.} \textbf{133}, 191002 (2024). https://doi.org/10.1103/PhysRevLett.133.191002

\bibitem{PandaX_CEvNS}
Z. Bo, et al. (PandaX Collaboration), First Indication of Solar $^{8}\mathrm{B}$ Neutrinos through Coherent Elastic Neutrino-Nucleus Scattering in PandaX-4T, \emph{Phys. Rev. Lett.} \textbf{133}, 191001 (2024). https://doi.org/10.1103/PhysRevLett.133.191001

\bibitem{LZ_CEvNS}
DS Akerib, et al., Searches for Light Dark Matter and Evidence of Coherent Elastic Neutrino-Nucleus Scattering of Solar Neutrinos with the LUX-ZEPLIN (LZ) Experiment, \emph{arXiv preprint arXiv:2512.08065} (2025).

\bibitem{cai2024reactor}
C. Cai, et al., Reactor neutrino liquid xenon coherent elastic scattering experiment, \emph{Phys. Rev. D} \textbf{110}, 072011 (2024). https://doi.org/10.1103/PhysRevD.110.072011

\bibitem{yu2026projection}
J. Yu, et al., Projection of purification performance for the RELICS experiment, \emph{arXiv preprint arXiv:2604.12572v2} (2026).

\bibitem{xie2026development}
L. Xie, et al., Development of a dual-phase xenon time projection chamber prototype for the RELICS experiment, \emph{Eur. Phys. J. C} \textbf{86}, 348 (2026).

\bibitem{duncan2005permeation}
B. Duncan, J. Urquhart, and S. Roberts, Review of measurement and modelling of permeation and diffusion in polymers, \emph{NPL Report DEPC MPR 012}, National Physical Laboratory, Teddington, UK (2005).

\bibitem{Sammartano_P35}
Sergio Giacomo Sammartano, Outgassing rates of PEEK, Kapton® and Vespel® polymers,  (2020).

\bibitem{OHanlon_Vacuum_Tech}
J. F. O'Hanlon, \emph{A User's Guide to Vacuum Technology}, 57-78 (2003). https://doi.org/10.1002/0471467162.ch4

\bibitem{chiggiato2020outgassing}
Paolo Chiggiato, Outgassing properties of vacuum materials for particle accelerators, \emph{arXiv preprint arXiv:2006.07124} (2020).

\bibitem{Huang2016outgassing}
A. Y. Huang, G. N. Kastanas, L. Kramer, C. E. Soares, and R. R. Mikatarian, Materials outgassing rate decay in vacuum at isothermal conditions, \emph{Proc. SPIE} \textbf{9952}, 995206 (2016).

\bibitem{Koros1981Oxygen}
W. J. Koros, J. Wang, and R. M. Felder,
Oxygen permeation through FEP Teflon and Kapton polyimide,
\emph{J. Appl. Polym. Sci.} \textbf{26}, 2805--2809 (1981).
doi: 10.1002/app.1981.070260832.

\bibitem{david1992peek}
L. David and S. Etienne,
Molecular mobility in para-substituted polyaryls. 1. Sub-$T_{\mathrm{g}}$
relaxation phenomena in poly(aryl ether ether ketone),
Macromolecules \textbf{25}, 4302--4308 (1992).

\bibitem{leonardi2013peek}
A. Leonardi, E. Dantras, J. Dandurand, and C. Lacabanne,
Dielectric relaxations in PEEK by combined dynamic dielectric
spectroscopy and thermally stimulated current,
J. Therm. Anal. Calorim. \textbf{111}, 807--814 (2013).

\bibitem{Grinham2017Outgassing}
R. Grinham and A. Chew,
A review of outgassing and methods for its reduction,
Appl. Sci. Converg. Technol. \textbf{26}, 95--109 (2017).

\bibitem{NASA_Bakeout}
NASA Small Spacecraft Systems Virtual Institute,
Bake Out, SSRI Knowledge Base, Integration and Test.

\bibitem{Sammartano2020Polymers}
S. G. Sammartano, I. Wevers, G. Bregliozzi, and P. Chiggiato,
Outgassing rates of PEEK, Kapton\textsuperscript{\textregistered}
and Vespel\textsuperscript{\textregistered} foils,
CERN-ACC-NOTE-2020-0039 (2020).

\end{thebibliography}
\end{document}